\documentclass[12pt,a4paper]{article}

\usepackage{a4wide}
\usepackage{amsmath}
\usepackage{bm}
\usepackage{amssymb}
\usepackage{hyperref}
\usepackage{epic}

\newcommand{\cN}{{\mathcal N}}

\newcommand{\cA}{C_A}
\newcommand{\zt}{\zeta_3}
\newcommand{\zfr}{\zeta_4}
\newcommand{\zf}{\zeta_5}

\newcommand{\dff}{d_{44}}

\newcommand{\DR}{\mathrm{DR}}

\begin{document}
\thispagestyle{empty}

\begin{center}
{\Large{\bf
Vanishing of the four-loop charge renormalization function in
$\cN=4$ SYM theory
}}
\vspace{15mm}

{\sc
V.~N.~Velizhanin}\\[5mm]

{\it Theoretical Physics Department\\
Petersburg Nuclear Physics Institute\\
Orlova Roscha, Gatchina\\
188300 St.~Petersburg, Russia}\\[5mm]

\textbf{Abstract}\\[2mm]
\end{center}

\noindent{We calculate the renormalization constants
of the maximally extended $\cN=4$ supersymmetric Yang-Mills theories
in the dimensional reduction scheme up to four loops.
We have found, that the beta-function is zero both from gauge and Yukawa vertices.
}
\newpage

\setcounter{page}{1}

30-years ago the three-loop calculations of the $\beta$-function in maximally extended $\cN=4$ supersymmetric Yang-Mills (SYM) theory were
performed~\cite{Avdeev:1980bh,Grisaru:1980nk,Caswell:1980yi}.
The result of calculations was zero as in one- and two-loop orders~\cite{Jones:1977zr,Poggio:1977ma}. Generalization to all loop orders was performed further in Refs.~\cite{Brink:1982wv,Howe:1983sr}.
A distinctive feature of these calculations was the necessity of a regularization that preserve supersymmetry. The calculations in Refs.~\cite{Grisaru:1980nk,Caswell:1980yi} were performed in superfield formalism, while in Ref.~\cite{Avdeev:1980bh} component approach was used. A calculation in components is conventional and can be easily done if one already has appropriate method for calculations in usual gauge theories, such as Quantum Chromodynamics (QCD). Namely in this way the calculations of the three-loop $\beta$-function in QCD~\cite{Tarasov:1980au} were extended to $\cN$=4 SYM theory in Ref.~\cite{Avdeev:1980bh}. However, the dimensional regularization, which was used for multi-loop calculations in QCD, violate supersymmetry because in supersymmetric theories one should keep the number of components of all spinors fixed. To restore supersymmetry one should add to the $4-2\epsilon$ gauge fields $2\epsilon$ scalar fields~\cite{Siegel:1979wq,Capper:1979ns}. So, in the calculation in $\cN=4$ SYM theory the Dimensional Reduction ($\DR$) scheme  prescribes to work with Dirac matrices in four dimensions, while the number of scalar and pseudoscalar fields should be equal $3+\epsilon$ rather then $3$. In this way the vanishing three-loop $\beta$-function was obtained both from gauge and Yukawa vertices~\cite{Avdeev:1980bh}. However, $\DR$ scheme contains internal contradictions~\cite{Siegel:1980qs}, which lead to the incorrect result in the higher-loop orders. Such example was found with the generalization of three-loop calculations to the $\cN=1$ and $\cN=2$ SYM theories~\cite{Avdeev:1982np}. It was found, that the $\beta$-functions of the gauge and Yukawa vertices do not
coincide starting from three loops\footnote{Despite the fact that the recent calculations give the same $\beta$-function from the gauge and Yukawa vertices in the case of $\cN=1$ SYM theory~\cite{Harlander:2006xq,Velizhanin:2008rw}, $\DR$ scheme still gives the different results in the case of $\cN=2$ SYM theory~\cite{Velizhanin:2008rw}}. Then, the investigations of applicability of $\DR$ scheme were performed~\cite{Avdeev:1981vf,Avdeev:1982np,Avdeev:1982xy} and estimations give, that for the $\cN=4$ SYM theory $\DR$ scheme should work up to five loops for the propagator type diagrams (see Table 1 in Ref.~\cite{Avdeev:1982np}).

The aim of the work presented in this paper is the calculation of the $\beta$-function in $\cN=4$ SYM theory from the gauge and Yukawa vertices in the framework of $\DR$ scheme to check the correctness of this scheme in the fourth order of perturbation theory. Moreover, our result for the four-loop renormalization constants can be used for the possible calculations at five-loop order in $\cN=4$ SYM theory,
such as calculation of anomalous dimension of Konishi operator~\cite{Konishi:1983hf}\footnote{In spite of anomalous dimensions of the twist-2 operators (including Konishi) are known
at five-loop order~\cite{Bajnok:2009vm,Lukowski:2009ce} (and at six-loop order for the most simple twist-3 operators~\cite{Velizhanin:2010cm}), direct calculations are performed only at four loops~\cite{Fiamberti:2007rj,Fiamberti:2008sh,Velizhanin:2009zz} (for twist-3 at five loops~\cite{Fiamberti:2009jw}).}
for the testing of integrability in the framework of AdS/CFT-correspondence.

Renormalization constants within MS-like schemes do not
depend on dimensional parameters (masses, momenta)~\cite{Collins:1974bg} and
have the following structure:
\begin{equation}
Z
=
1+\sum^\infty_{n=1}z^{(n)}\!\!\left(\alpha,g^2\right)\epsilon^{-n},\label{Z}
\end{equation}
where $\alpha$ is the gauge fixing parameter.
The renormalization constants define corresponding anomalous dimensions:
\begin{equation}
\gamma(\alpha,g^2)=
g^2\frac{\partial}{\partial g^2}\ z^{(1)}\!(\alpha,g^2)=\sum^\infty_{n=1}g^{2(n+1)}\gamma^{(n)}.\label{defga}
\end{equation}
The renormalization of the coupling constant is related with the renormalization constant
of the corresponding vertex and the  renormalization constants of the fields entering into this vertex.
For the triple vertices we have
\begin{equation}
Z_{g^2}=Z_{jjk}^2 Z_j^{-2} Z_k^{-1}\,,\label{gbz}
\end{equation}
where $Z_{jjk}$ and $Z_j$ are the renormalization constants for the triple vertices and
the wave functions correspondingly.
From the last equation one obtains the charge renormalization $\beta$-function as
\begin{equation}
\beta_{jjk}\left(g^2\right)=
g^2 \left[2\,\gamma_{jjk}\!\left(\alpha,g^2\right)-2\,\gamma_{j}\!\left(\alpha,g^2\right)
-\gamma_{k}\!\left(\alpha,g^2\right)\right].\label{beta}
\end{equation}
We have calculated the renormalization constants for all fields and for the ghost-ghost-gluon and fermion-fermion-scalar vertices, that give us $\beta$-function from the two different type of vertices.

The calculations of the renormalization constants within MS-like scheme can
be reduced to the calculation only
of massless propagator type diagrams by means of the method of
infrared rearrangement~\cite{Vladimirov:1979zm}.
In the case of the gauge (ghost-ghost-gluon in our case) or Yukawa (fermion-fermion-scalar) vertices it means that we can nullify the momentum of the external gauge or scalar fields, correspondingly,
reducing the calculation of the $Z_{jjk}$ to the propagator type diagrams.

Calculations of the renormalization constants were made with our program BAMBA~\cite{BAMBA} based on the algorithm of Laporta~\cite{Laporta:2001dd}
(see also~\cite{Misiak:1994zw,Chetyrkin:1997fm,Czakon:2004bu}), which we used in our previous calculations~\cite{Velizhanin:2009zz,Velizhanin:2009gv}.
All calculations were performed with FORM~\cite{Vermaseren:2000nd},
using FORM package COLOR~\cite{vanRitbergen:1998pn} for evaluation of the color traces
and with the Feynman rules from Ref.~\cite{Gliozzi:1976qd}.
For the dealing with a huge number of diagrams we use a program DIANA~\cite{Tentyukov:1999is},
which call QGRAF~\cite{Nogueira:1991ex} to generate all diagrams enumerated in Table~\ref{numbdiagr}.
\begin{table}[ht]
\centering
\vskip 2mm
\begin{tabular}{|c|c|c|c|c|}
  \hline
    & 1-loop & 2-loop & 3-loop & 4-loop \\
    \hline
Ghost wave function    & 1 & 8 & 158 & 4.563 \\
  \hline
Scalar wave function    & 2 & 34 & 930 & 37.014 \\
  \hline
Fermion wave function    & 3 & 40 & 1.210 & 51.465 \\
  \hline
Gluon wave function   & 5 & 58 & 1.513 & 57.664 \\
  \hline
Ghost-ghost-gluon vertex & 2  & 47  & 1.462& 57.939 \\
  \hline
Fermion-fermion-scalar vertex & 5  & 183& 8.845 & 517.576 \\
  \hline
  \hline
Sum & 18  & 370 & 14.118& 726.221\\
  \hline
\end{tabular}
\caption{The number of diagrams for calculations up to four-loop order.}
\label{numbdiagr}
\end{table}

The results of calculations up to four-loop order are the following:
\begin{eqnarray}
\gamma_{3}& =&
          - 2\,\cA\,a
          + \,\cA^2\,a^2
         +\left(- \frac{59}{16}
          - \frac{63}{4}\,\zt\right)\cA^3\,a^3
\nonumber\\
&+&\left(\!
\left(
          - \frac{305}{192}
          - \frac{16325}{96}\,\zf
          + \frac{45}{16}\,\zfr
          + \frac{2797}{96}\,\zt
\right)\cA^4
+\left(\!
          - 9
          + \frac{125}{4}\,\zf
          + \frac{185}{4}\,\zt
\right) \dff\!
\right)a^4,
\label{gammagluon}
\\
\tilde\gamma_{3}& =&
          \frac{1}{2}\cA\,a
          - \frac{5}{4}\,\cA^2\,a^2
          + \left(\frac{155}{32}
          + \frac{63}{8}\,\zt\right)\,\cA^3\,a^3
\nonumber\\
&+&\left(
\left(
            \frac{5849}{384}
          + \frac{14725}{192}\,\zf
          + \frac{81}{32}\,\zfr
          + \frac{499}{192}\,\zt
\right)\cA^4
+\left(
            \frac{9}{2}
          - \frac{265}{8}\,\zf
          - \frac{49}{8}\,\zt
\right) \dff
\right)a^4,
\label{gammaghost}
\\
\tilde\gamma_{1}& =&
          -\frac{1}{2} \cA\,a
          - \frac{3}{4}\,\cA^2\,a^2
          + 3\,\cA^3\,a^3
\nonumber\\
&+&\left(
\left(
            \frac{231}{16}
          - \frac{25}{3}\,\zf
          + \frac{63}{16}\,\zfr
          + \frac{103}{6}\,\zt
\right)\cA^4
+\left(
          - \frac{35}{2}\,\zf
          + 17\,\zt
\right) \dff
\right)a^4,
\label{gammaghostghostgluon}
\\
\gamma_{\phi}& =&
          - 2\,\cA\,a
          - \cA^2\,a^2
+\left(     \frac{23}{4}
          - \frac{27}{2}\,\zt
\right)\cA^3\,a^3
\nonumber\\
&+&\left(
\left(
            \frac{561}{16}
          - \frac{7145}{48}\,\zf
          + \frac{57}{8}\,\zfr
          + \frac{2597}{48}\,\zt
\right)\cA^4
+\left(   - 7
          + \frac{185}{2}\,\zf
          - \frac{59}{2}\,\zt
\right) \dff
\right)a^4,
\label{gammascalar}
\\
\gamma_{\lambda}& =&
          - 4\,\cA
          + 6\,\cA^2\,a^2
          +\left(- \frac{101}{4}
          - 27\,\zt\right)\,\cA^3\,a^3
\nonumber\\
&+&\left(
\left(
          - \frac{5591}{48}
          - \frac{3185}{12}\,\zf
          + \frac{51}{8}\,\zfr
          - \frac{1009}{24}\,\zt
\right)\cA^4
+\left(
          - 8
          - 320\,\zf
          + 142\,\zt
\right) \dff
\right)a^4,
\label{gammagluino}
\\
\gamma_{4}& =&
          - 5\,\cA\,a
          + \frac{11}{2}\,\cA^2\,a^2
          +\left(- \frac{179}{8}
          - \frac{135}{4}\,\zt\right)\,\cA^3\,a^3
\nonumber\\
\hspace{-1mm}&\hspace{-9mm}-&\hspace{-5mm}\left(\!
\left(
            \frac{9499}{96}
          + \frac{10875}{32}\,\zf
          - \frac{159}{16}\,\zfr
          + \frac{1439}{96}\,\zt
\right)\cA^4
+\!\left(
            \frac{23}{2}
          + \frac{1095}{4}\,\zf
          - \frac{509}{4}\,\zt
\right)\!\dff\!
\right)a^4\!,
\label{gammagluinogluinoscalar}
\end{eqnarray}
with $a={g^2}/{(4\pi)^2}$ and the following Casimir operators of gauge group $SU(N)$: $C_A=N$, $d_{44} =N^2(N^2+36)/24$. 
Here $\tilde\gamma_1$ and $\gamma_4$ are the anomalous dimensions of the ghost-ghost-gluon and fermion-fermion-scalar vertices, and $\gamma_3$, $\tilde\gamma_3$, $\gamma_\phi$ and $\gamma_\lambda$ are those of gluon, ghost, scalar, and fermion fields, respectively.
The three-loop results for the same quantities can be found in Ref.~\cite{Avdeev:1980bh}.

Substituting the obtained $\gamma$-functions into Eq.~(\ref{beta})
we have found both from the ghost-ghost-gluon and fermion-fermion-scalar that the $\beta$-function is equal to zero:
\begin{equation}\label{betag}
\beta^{4-\mathrm{loop}}(a)\ =\ 0.
\end{equation}

So, we have found, that the gauge and Yukawa couplings are renormalized in the same way in
$\cN=4$ SYM theory. Hence, the $\DR$-scheme
preserves supersymmetry and works correctly in these model up to four loops.
To conclude, we note again that the obtained renormalization constants can be used for the possible calculation of the
five-loop anomalous dimension of Konishi operator in $\cN=4$ SYM theory.

\subsection*{Acknowledgements}
We would like to thank K.G. Chetyrkin, L.N. Lipatov, A.I. Onishchenko, A.V. Smirnov, V.A. Smirnov and A.A. Vladimirov
for useful discussions.
This work is supported by
RFBR grants 10-02-01338-a, RSGSS-65751.2010.2.


\end{document}